\documentclass[11pt]{article}

\usepackage{amsmath, amssymb}

\usepackage{graphicx}

\usepackage{geometry}
\usepackage[numbers]{natbib}

\usepackage[hyperfootnotes=false, colorlinks=true, linkcolor=blue, citecolor=blue, urlcolor=blue]{hyperref} 
\usepackage[nameinlink]{cleveref} 

\usepackage{xurl}

\newcommand{\figref}[1]{\cref{#1}}
\newcommand{\eqreftext}[1]{eq.~\eqref{#1}}

\title{\Large\itshape Correlation Between Nighttime Light and Various Statistical Indicators \\in Japan}
\author{Shoichi Otomo}
\date{}

\begin{document}

\maketitle

\begin{abstract}

This manuscript is an English translation and extended version of a paper originally published in Japanese~\cite{otomo}.
Recent years have seen a rapid expansion in applying satellite remote sensing and big data to economic analysis. 
In particular, satellite-observed nighttime light intensity has proven to be closely correlated with key socio-economic indicators, including gross domestic product, employment rates, population density, and educational attainment across countries. 
In this paper, I first outline the methodology for calculating regional nightlight intensity across Japanese prefectures and municipalities. 
Next, to evaluate the versatility of nightlight data, I examine the relationship between city-level nightlight intensity and various public socio-economic metrics, using Japan as a case study. 
The findings suggest that nightlight intensity can serve as a effective proxy for these diverse indicators. 
Finally, I explore the potential of nightlight data for enabling rapid economic impact assessments during sudden social disruptions, such as the COVID-19 pandemic.

\end{abstract}


\section{Introduction}
\subsection{Background}
Advances in information and communication technology (ICT) and geospatial data analysis have vastly expanded our capacity to process large-scale spatial datasets (big data). This technological shift has established new frameworks for extracting socio-economic value from spatial observation. Simultaneously, national policies worldwide have increasingly encouraged open data initiatives, facilitating the secondary usage of public administrative records.

In parallel, satellite observation technology has undergone rapid evolution, enabling the continuous acquisition of high-resolution, high-frequency Earth observation data. While early efforts to capture aerial perspectives for economic research in the late 19th century relied on basic platforms such as balloons, kites, and pigeons, the field of remote sensing has shifted dramatically over the past decade \cite{donaldson}. Today, petabyte-scale satellite archives are publicly accessible, serving interdisciplinary domains spanning economics, geography, and urban engineering.

Satellite data now provide frequent and granular observations used across diverse applications, including crop yield modeling, urban footprint mapping, infrastructure management, environmental pollution monitoring, and disaster tracking. Among these resources, nighttime light (NTL) imagery has emerged as a particularly versatile dataset, serving as an effective proxy for human activities and economic development.

\subsection{Related Works}
Nighttime light intensity has been widely validated as a reliable proxy for tracking macroeconomic dynamics. Henderson et al. (2012) \cite{henderson} demonstrated that NTL fluctuations closely track GDP variations during major economic and social disruptions, such as the Asian Financial Crisis in Indonesia, the Rwandan genocide, and mining booms in Madagascar. They emphasized that standard national accounts (e.g., official GDP) suffer from measurement errors, reporting delays, and cross-country comparability issues—particularly in developing contexts. Under such constraints, satellite-observed nightlight offer an objective, consistent framework for measuring local and national economic activity.

In low-income and regional settings, Kurata (2017) \cite{kurata} analyzed administrative data in Bangladesh, demonstrating that NTL radiance correlates significantly not only with primary demographic and infrastructure metrics (such as population density and road access) but also with broader welfare indicators, including poverty rates, literacy, and child health outcomes.

\section{Objective}
\begin{figure}
    \centering
    \includegraphics[width=130mm]{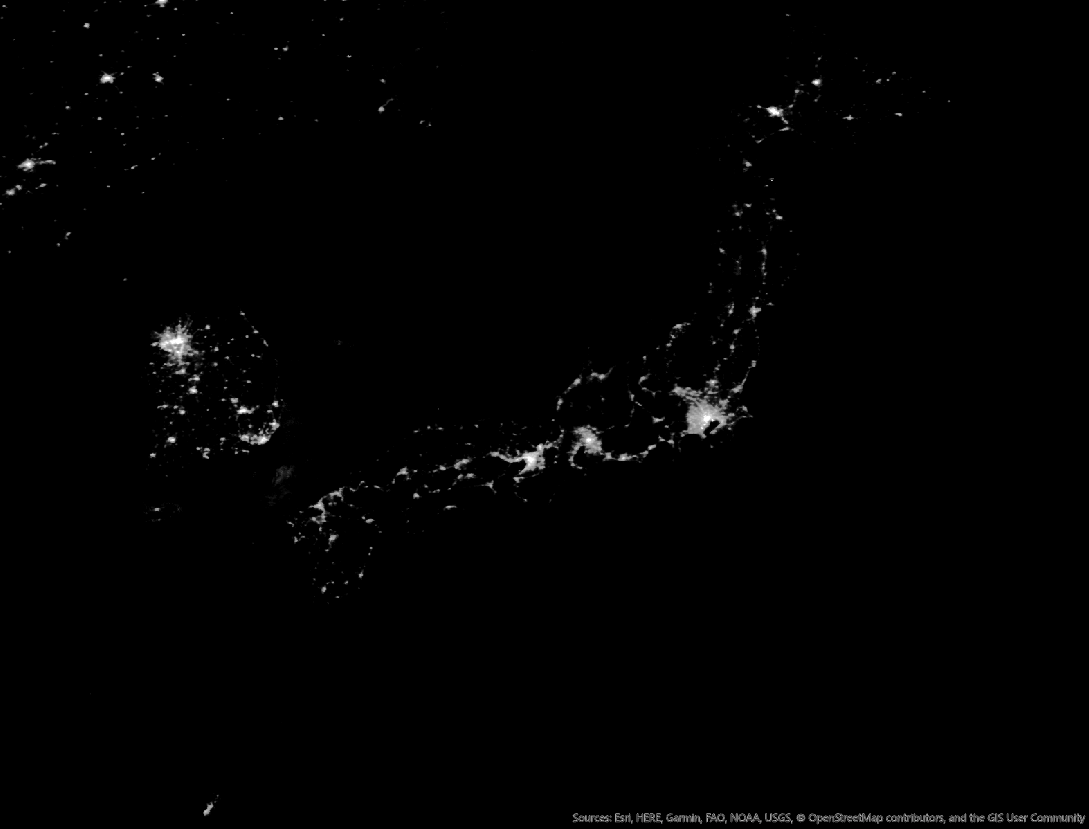}
    \caption{NightLight around Japan}
    \label{fig:around_jpnNLlight}
\end{figure}

Compared to economic data from developing countries, data published by the Japanese government is generally considered to be highly accurate. On the other hand, much of the data released by the Japanese government requires an enormous amount of time for aggregation, making it difficult to claim that it is suitable for conducting economic forecasting or analysis that reflects sudden social events.

Furthermore, unlike the cases of developing countries, nighttime lights in Japan are expected to possess unique characteristics. These characteristics include associations with economic and social conditions, encompassing both synchronized trends and time-series correlations.

Therefore, this paper analyzes the relationship between government data and nighttime lights within Japan. For the nighttime lights, following Henderson et al.(2012) \cite{henderson}, I utilize data obtained by the Defense Meteorological Satellite Program(DMSP), a weather satellite of the U.S. Air Force.

\section{Data Construction}
\subsection{Nighttime Lights Data Utilized}

From the nighttime lights data published by the National Oceanic and Atmospheric Administration(NOAA), radiance data that has undergone cloud correction in advance is processed and distributed as annual average values from March 1996 to July 2011. In this paper, among the distributed datasets, I utilize the data from January 2010 to December 2010, which is available as an annual average value. The data is distributed in TIFF format and covers the entire globe. From this global dataset, the area around Japan, which is the subject of analysis, is extracted \figref{fig:around_jpnNLlight}.

\subsection{Conversion of Nighttime Lights Data into a Shapefile}
\begin{figure}
    \centering
    \includegraphics[width=130mm]{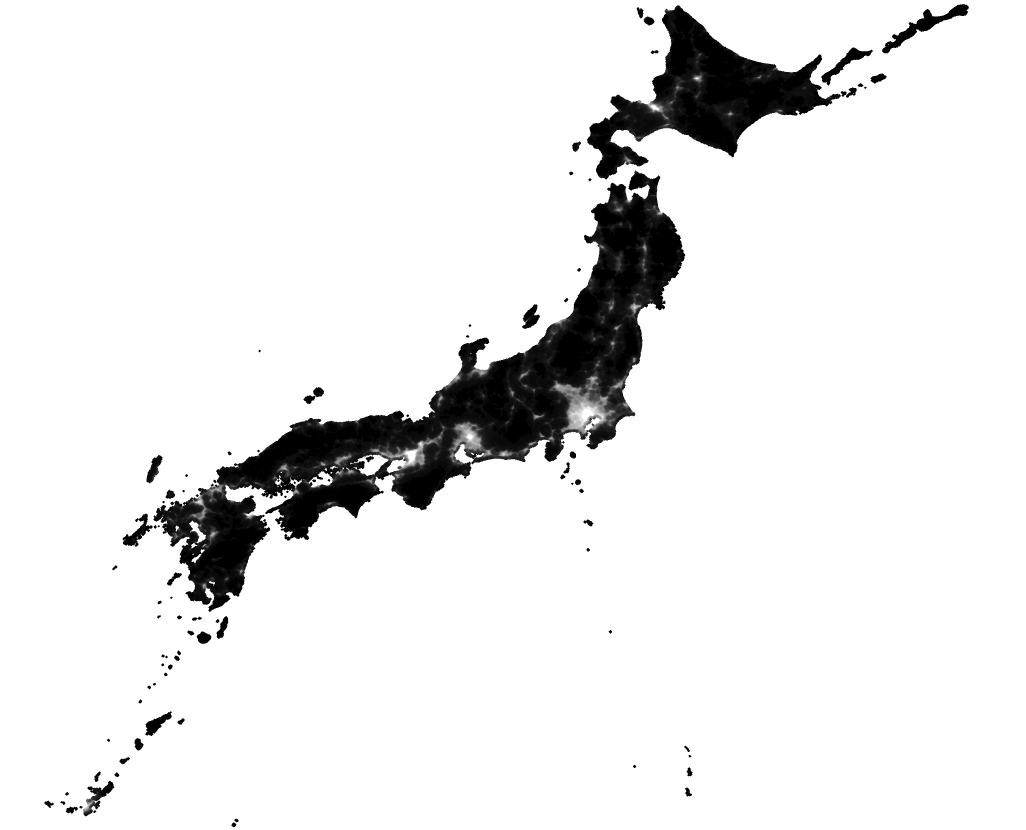}
    \caption{Night Light Of Japan}
    \label{fig:JPN_NL1}
\end{figure}

To calculate nighttime lights using a GIS tool, a shapefile of a point grid generated at 500 meter intervals is created. Subsequently, the nighttime light values corresponding to the latitudes and longitudes of the aforementioned distributed data are appended to this grid. By clipping this dataset to the Japanese territory using the "National Land Numerical Information(Administrative Districts)" provided by the Ministry of Land, Infrastructure, Transport and Tourism(MLIT), a shapefile featuring the intensity of nighttime lights in Japan as attribute values is generated \figref{fig:JPN_NL1}.

\subsection{Calculation of Nighttime Lights Data at the Municipal Level}
\begin{figure}
    \centering
    \includegraphics[width=130mm]{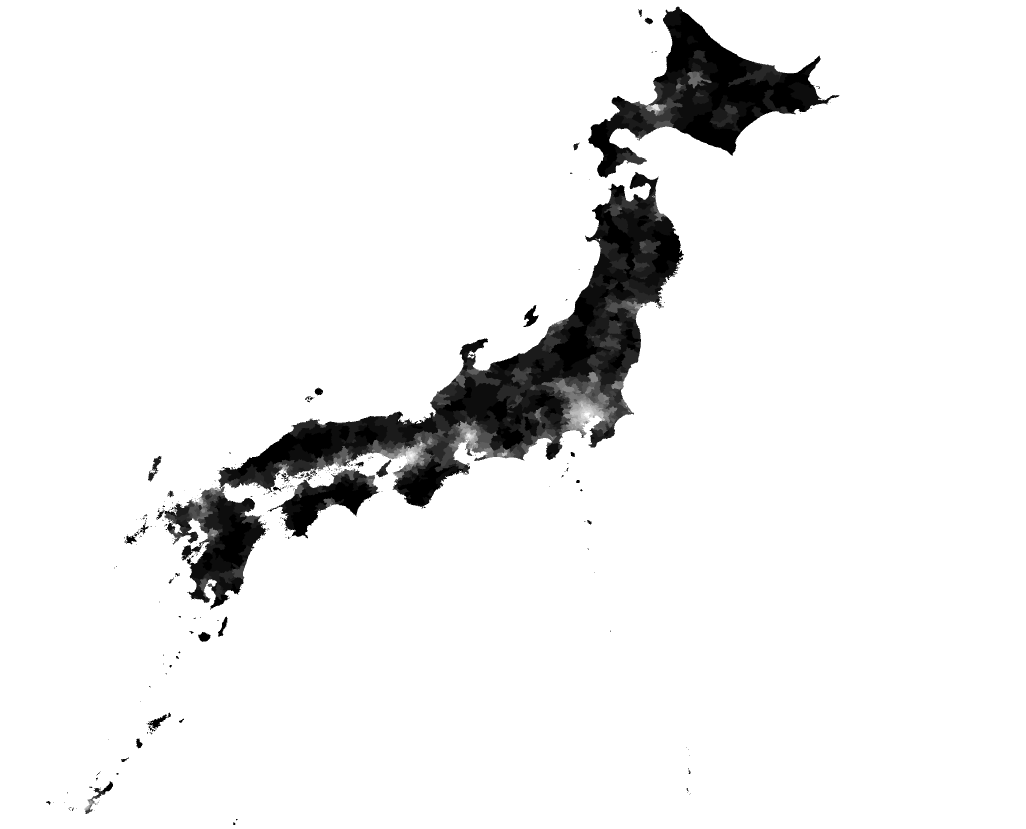}
    \caption{Average Night Light Intensity by Municipality}
    \label{fig:JPN_NL2}
\end{figure}

\begin{figure}
    \centering
    \includegraphics[width=130mm]{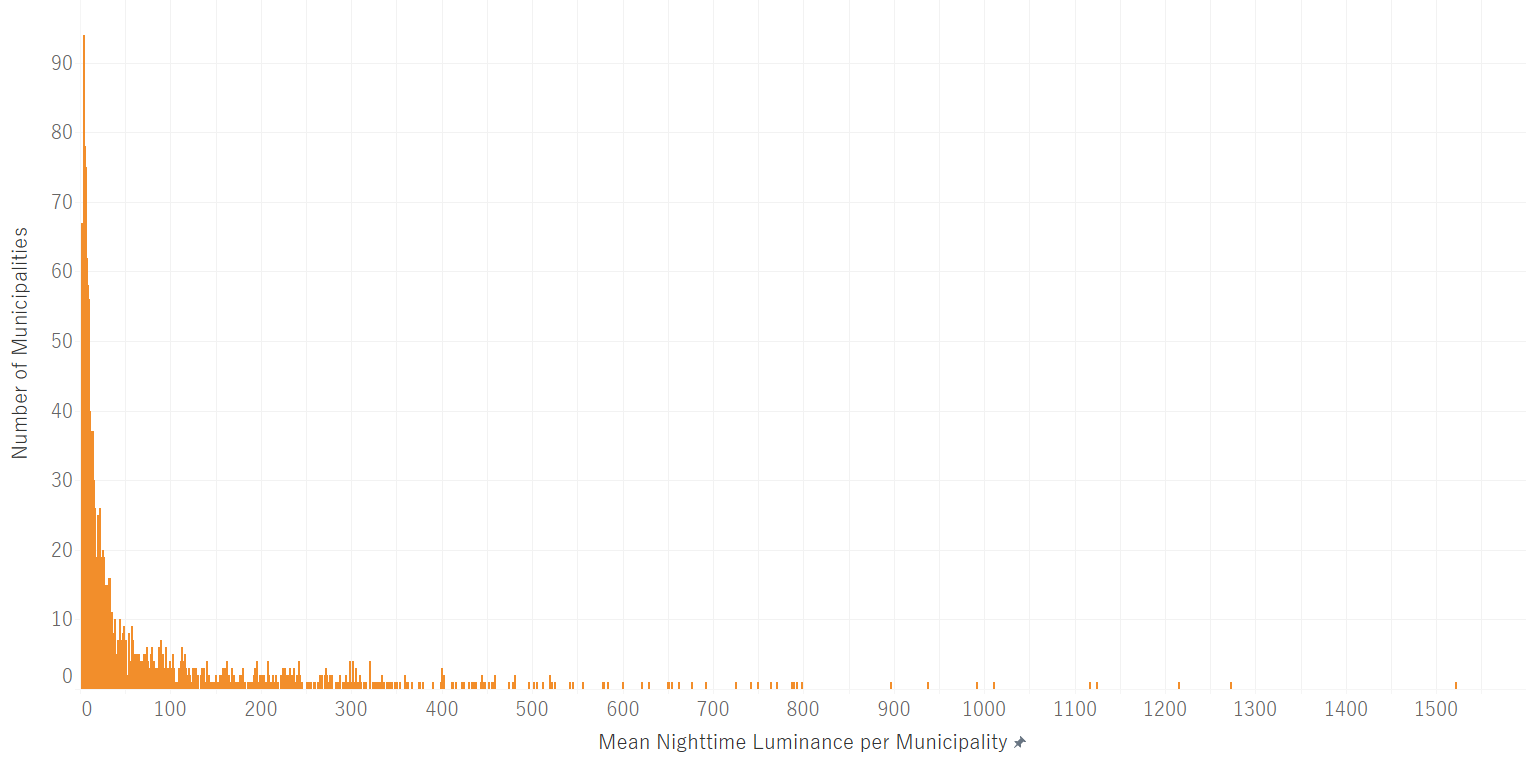}
    \caption{Average Histogram of Nighttime Luminance by Municipality}
    \label{fig:JPN_NL_hist}
\end{figure}

Many statistical datasets and censuses utilize municipalities as the unit of aggregation. To calculate the associations with these various statistical data, the nighttime lights data created in the previous section is spatially joined with the "System of Social and Demographic Statistics: Statistical Observations of Prefectures, Shi, Ku, Machi and Mura" provided by the Statistics Bureau of the Ministry of Internal Affairs and Communications(MIC). The condition for this spatial join is set to the average value of nighttime lights for each municipality \figref{fig:JPN_NL2}. Additionally, a histogram is constructed to examine the distribution of the average nighttime light values across municipalities \figref{fig:JPN_NL_hist}. As shown by this distribution, the brightness of nighttime lights by municipality is heavily skewed. Therefore, in some of the subsequent graphs, a logarithmic transformation is applied to the nighttime lights to facilitate a visual understanding.

\section{Analysis Results}
\subsubsection{Nighttime Light Intensity by Prefecture}

\begin{figure}
    \centering
    \includegraphics[width=130mm]{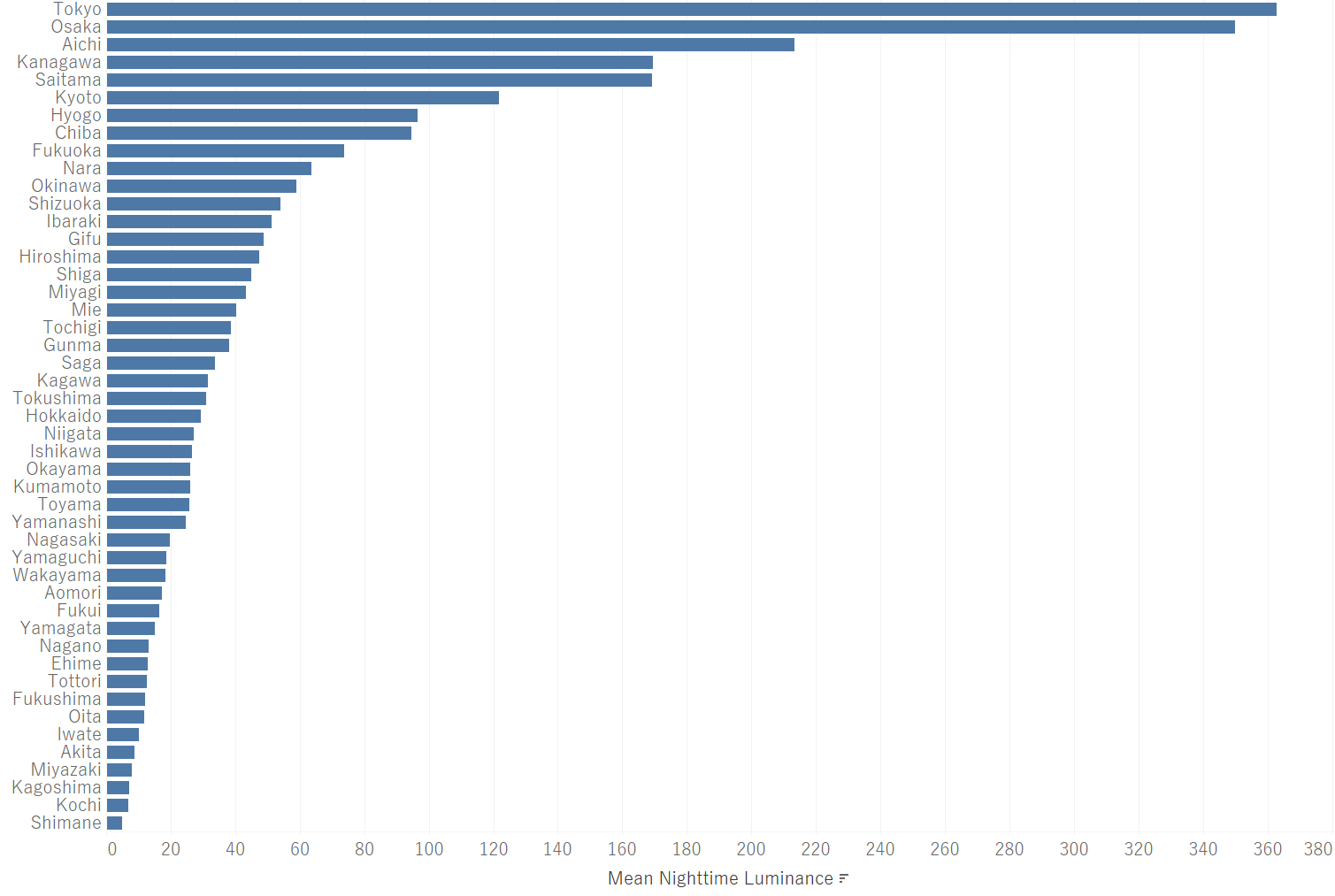}
    \caption{Average Nighttime Light Luminance by Prefecture}
    \label{fig:JPN_NL_mean_Pre}
\end{figure}

Nagae et al.(2018) \cite{nagae} state that urban centers can be defined from various perspectives, such as human activities, the concentration of buildings, and the spatial distribution of population. In this study, however, an urban center is specifically defined as an economically or commercially developed area.

Given this definition, it is evident from Figures 2 and 3 that urban centers exhibit greater brightness. Therefore, to first grasp the extent of nighttime light intensity in each prefecture, the average nighttime light values for each prefecture are arranged in descending order \figref{fig:JPN_NL_mean_Pre}.

\subsubsection{Nighttime Light Intensity by Municipality}
\begin{figure}
    \centering
    \includegraphics[width=130mm]{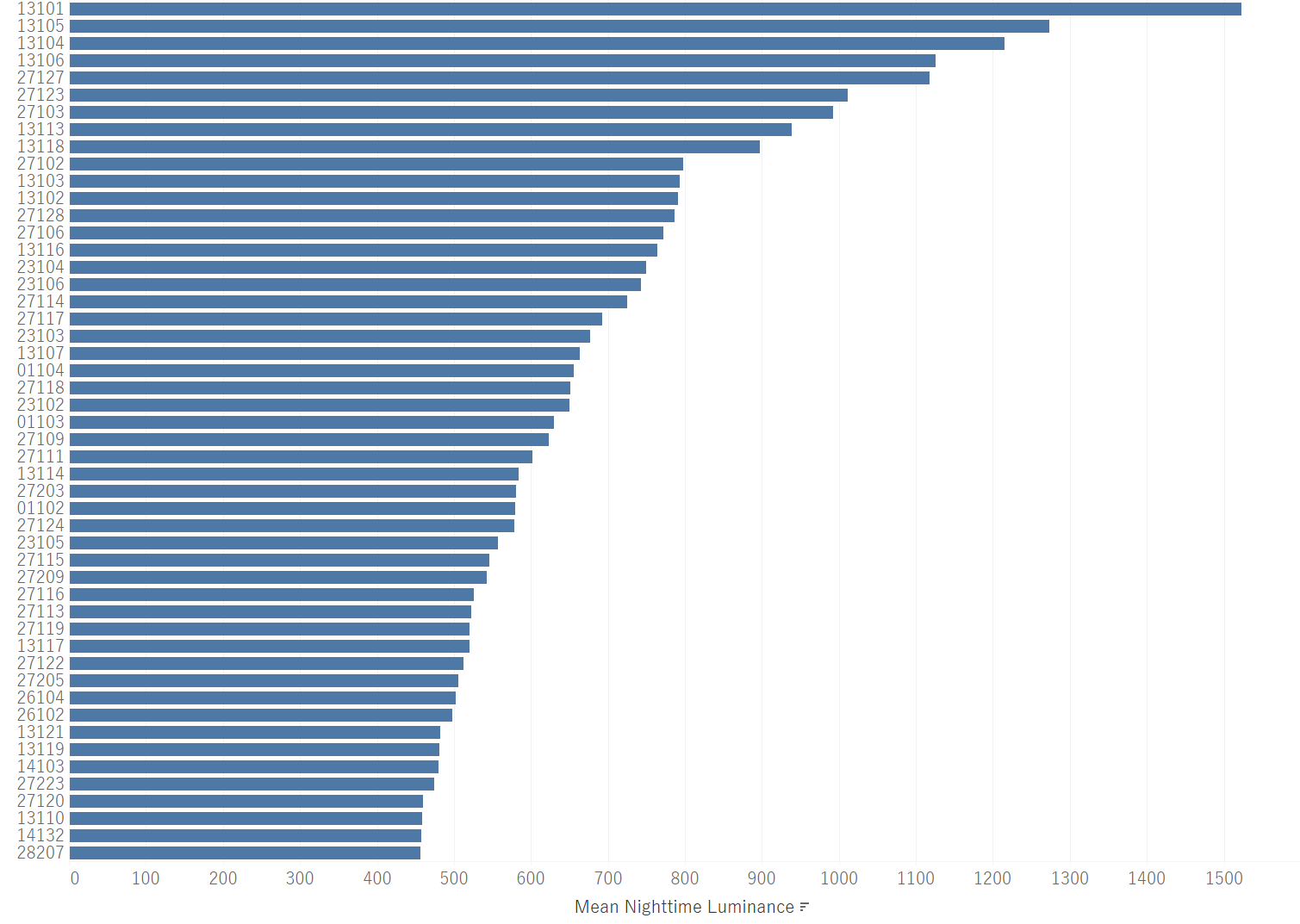}
    \caption{Average Nighttime Light Luminance by municipality}
    \label{fig:JPN_NL_mean_Muni}
\end{figure}

Next, within each prefecture, the data is arranged in order of nighttime light intensity by municipality. While this generally corresponds to the prefectural rankings, Figure 6 below shows the top 50 municipalities ranked by nighttime light intensity in descending order\figref{fig:JPN_NL_mean_Muni}.

\subsection{Correlation Between Nighttime Lights and Various Statistical Indicators by Municipality}
In this section, I examine the correlations and associations between nighttime lights and various statistical indicators.

\subsubsection{Relationship between Store Density and Nighttime Light Intensity in Each Municipality}
\begin{figure}
\begin{center}
\includegraphics[width=130mm]{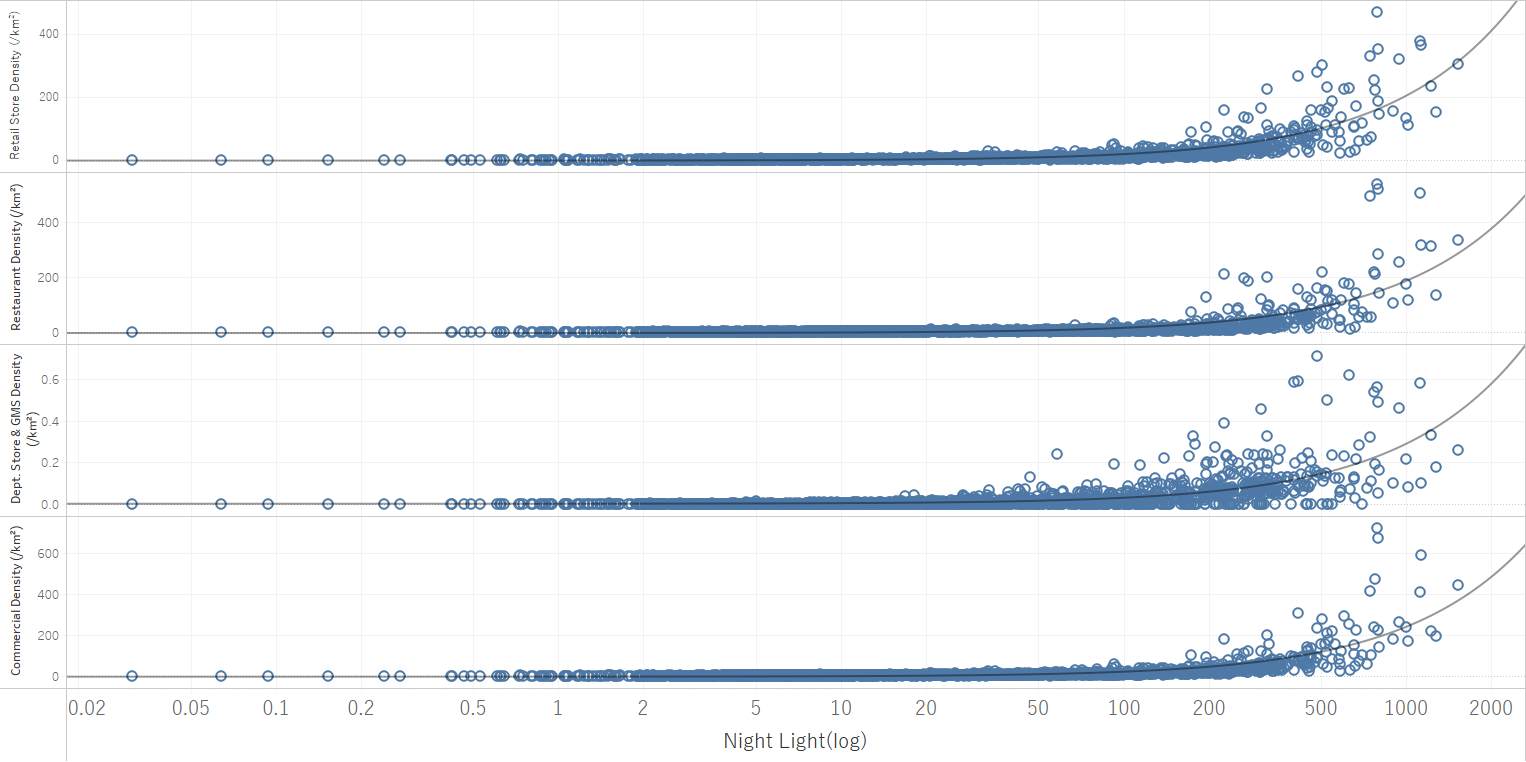}
\end{center}
\caption{Correlation between Nighttime Light Radiance and the Number of Stores}
\label{fig:JPN_NL_corr_tenpo}
\end{figure}

Retail stores, restaurants, commercial offices, department stores, and large supermarkets operate during the night and significantly affect nighttime light intensity; therefore, a high correlation coefficient is expected when conducting correlation analysis. At this time, instead of the number of stores($shop$), the store density($density$) within the area($area$) of each municipality($j$) was utilized, as shown in \figref{fig:JPN_NL_corr_tenpo}. The data is cited from the 2016 Economic Census included in the Starter Pack, a data package provided by ESRI Japan Corporation. Note that the store density was calculated using \eqreftext{eq:tenpo_density}, which yielded the results shown in \eqreftext{eq:density_corr}.

\begin{equation}
\label{eq:tenpo_density}
\begin{aligned}
 density_j = \sum_{i=1}^{n} \frac{shop_{i,j}}{area_j}
\end{aligned}
\end{equation}

As is evident from \figref{fig:JPN_NL_hist}, the average nighttime light values by municipality exhibit a heavily skewed distribution; therefore, a logarithmic transformation was applied. By performing a power-law approximation of the average nighttime light($L_j$) within each municipality($j$), the equations for retail stores($S$), restaurants($R$), commercial offices($F$), and department stores($D$) were obtained as shown in \eqreftext{eq:density_corr} below. Since $p\text{-}\mathit{value} < 0.0001$ in all cases, it can be argued that the explanatory power is remarkably high.

\begin{equation}
\label{eq:density_corr}
\begin{aligned}
S_j &= 0.153042L_j^{0.98}, \quad R^2 = 0.879967 \\
R_j &= 0.045149L_j^{1.11}, \quad R^2 = 0.856516 \\
F_j &= 0.148674L_j^{0.98}, \quad R^2 = 0.872701 \\
D_j &= 0.000376L_j^{0.96}, \quad R^2 = 0.799423
\end{aligned}
\end{equation}

\subsubsection{Relationship between Medical Services and Nighttime Light Intensity in Each Municipality}
Responses under the state of emergency associated with the COVID-19 pandemic were implemented at the prefectural level. Particularly in prefectures and major metropolitan areas, the number of infected individuals in each municipality became a frequent topic of discussion, which subsequently led to frequent media coverage regarding medical standards and the number of hospital beds in each prefecture.

However, medical care is generally considered a service where regional disparities are undesirable. According to Tanaka \cite{tanaka_med}, for the majority of medical care, long-term care up to a certain limit, and a certain portion of education and childcare for the next generation, a mechanism that allows anyone to access high-quality services is desirable. This is crucial not only for maintaining social stability but above all from the perspective of fundamental human dignity. Tanaka \cite{tanaka_med} also notes that public opinion surveys indicate the majority of the public shares this view.

\begin{figure}
\begin{center}
\includegraphics[width=130mm]{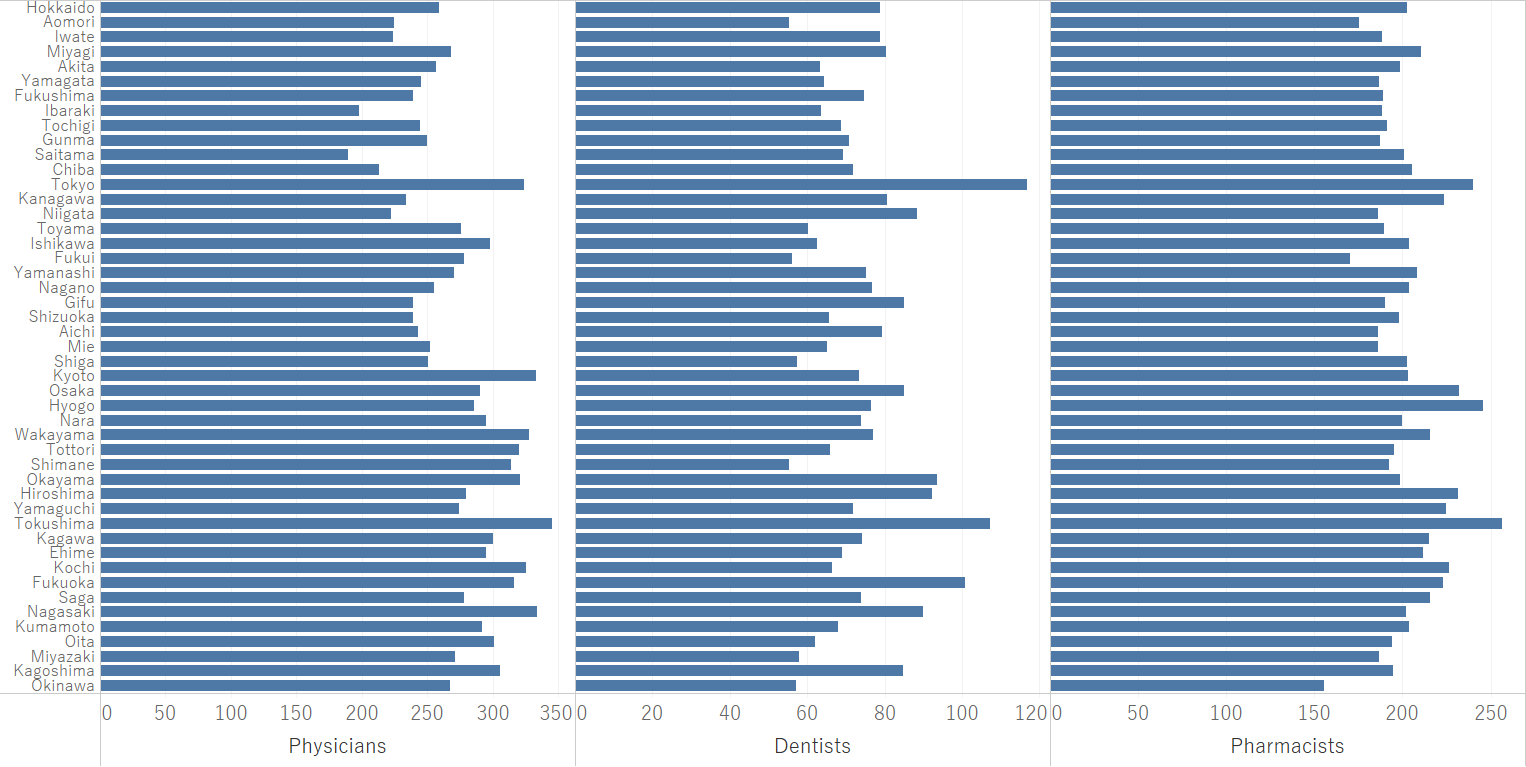}
\end{center}
\caption{Number of Physicians, Dentists, and Pharmacists per 100,000 Population}
\label{fig:med_staff_per10}
\end{figure}

Therefore, I refer to the "Number of Physicians, Dentists, and Pharmacists per 100,000 Population, by Prefecture, 16 Major Cities, and Core Cities, by Category of Facility/Activity, Sex, and Place of Employment" \cite{mhlw2008} provided by the Ministry of Health, Labour and Welfare, as shown in \figref{fig:med_staff_per10}. Although there are slight variations among prefectures, it can be observed that the number of practicing professionals remains relatively constant relative to the population size.

\begin{figure}[h]
\begin{center}
\includegraphics[width=130mm]{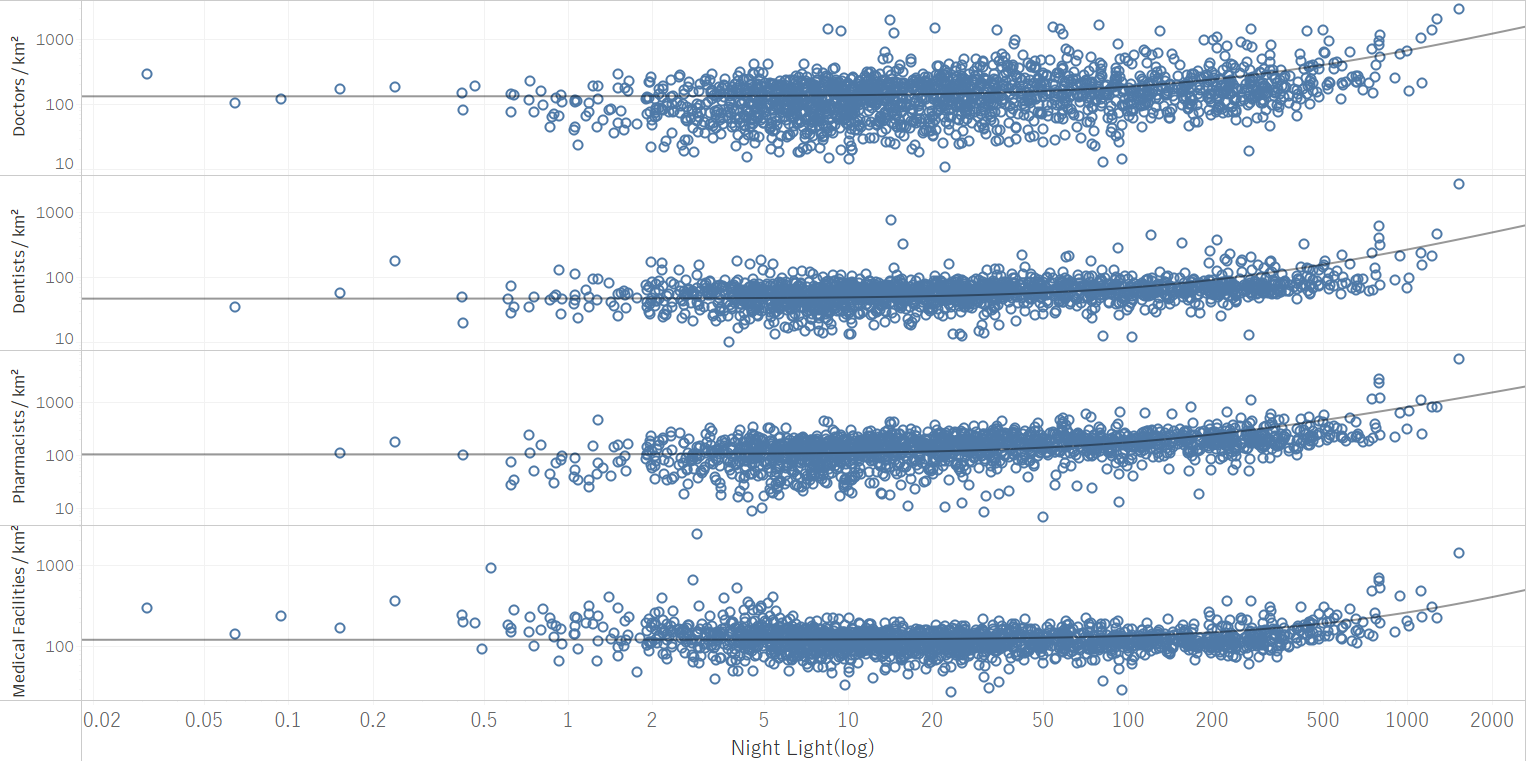}
\end{center}
\caption{Correlation between Nighttime Light Radiance(Municipal Average) and the Number of Physicians per 100,000 Population}
\label{fig:JPN_NL_corr_med}
\end{figure}

Next, I examine the correlation between the average nighttime lights of municipalities and the numbers of practicing physicians, dentists, pharmacists, and medical institutions in each municipality. Letting $Doc_j$, $Den_j$, $Pha_j$, and $Med_j$ represent the number of physicians, dentists, pharmacists, and medical institutions in each municipality($j$), respectively, I obtain \eqreftext{eq:med_corr} and \figref{fig:JPN_NL_corr_med} below.

\begin{equation}
\label{eq:med_corr}
\begin{aligned}
&Doc_j = 153.874L_j^{(0.07)}, R^2 = 0.0239618 \\
&Den_j = 45.7094L_j^{(0.10)}, R^2 = 0.125149 \\
&Pha_j = 108.759L_j^{(0.13)}, R^2 = 0.138838 \\
&Med_j = 116.893L_j^{(0.08)}, R^2 = 0.0159538
\end{aligned}
\end{equation}

From \figref{fig:med_staff_per10}, \figref{fig:JPN_NL_corr_med}, and \eqreftext{eq:med_corr}, it can be inferred that regional disparities in medical services within Japan are minimal. Furthermore, unlike commercial stores, it has become clear that there is virtually no correlation between medical services and nighttime lights.

\subsubsection{Relationship between Non-Flush Toilet Population and Nighttime Light Intensity in Each Municipality}

\begin{figure}[h]
\begin{center}
\includegraphics[width=130mm]{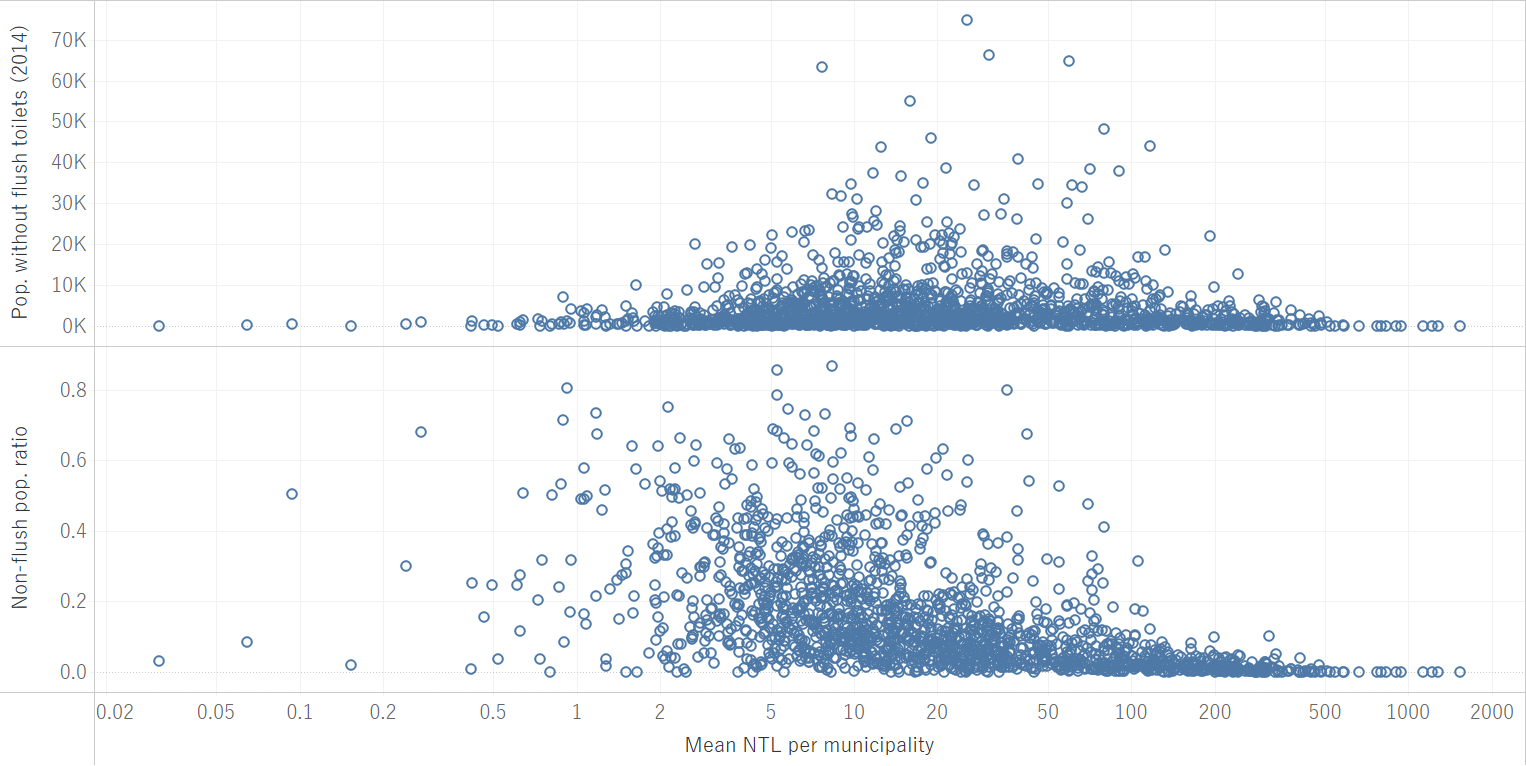}
\end{center}
\caption{Correlation between Nighttime Lights and Non-Flush Toilet Population Percentage}
\label{fig:washlet_popl}
\end{figure}

When utilizing the non-flush toilet rate as a basic social indicator for infrastructure, it is desirable that the non-flush toilet population and its percentage within each municipality show no regional disparities, similar to the case of medical services. Furthermore, although there is a certain correlation with brightness, this correlation is extremely low. As illustrated in \figref{fig:washlet_popl}, the non-flush toilet population percentage is less than 1\% at maximum and asymptotically approaches zero as nighttime light brightness increases. For instance, according to the glossary published by Kamakura City \cite{kamakura}, a flush toilet system is defined as a system where human waste can be treated via public sewerage, septic tanks, community plants, etc., making flush toilets available, whereas other methods such as vault toilets are classified as non-flush. Based on this, it can be argued that there is almost no regional disparity regarding the non-flush toilet population as well.

\subsubsection{Relationship between Average income and Nighttime Light Intensity in Each Municipality}

\begin{figure}
\begin{center}
\includegraphics[width=130mm]{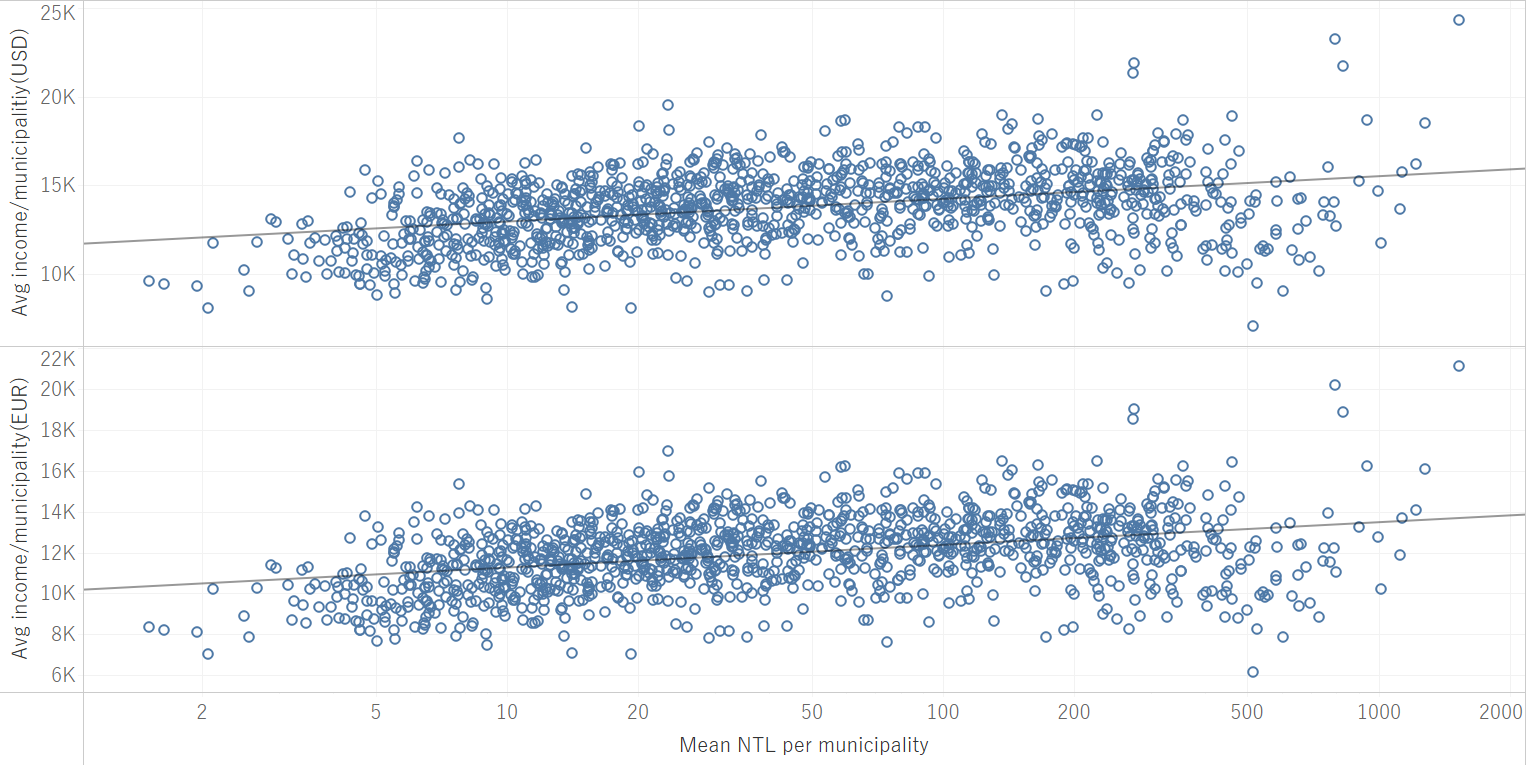}
\end{center}
\caption{Correlation between Nighttime Light Intensity and Average Income by Municipality}
\label{fig:income_nl_graph}
\end{figure}

Using the average income in each municipality, I examine the correlation between nighttime light intensity and income, as shown in \figref{fig:income_nl_graph}. Data for average income were obtained from the 2013 Housing and Land Survey \cite{estat_tochi} conducted by the Statistics Bureau of the Ministry of Internal Affairs and Communications.

As for the calculation method, the median value of each annual income bracket (in 10,000 JPY) was used for the household annual income categories. For the highest bracket of 15~million JPY or more, the value was set to 20~million JPY, defined as shown in \eqref{eq:wedge_class}.

\begin{equation}
\label{eq:wedge_class}
\mathit{income}_i = \{50, 150, 250, 350, 450, 600, 850, 1250, 2000\} \quad (\text{in 10,000 JPY})
\end{equation}

Next, as expressed in \eqref{eq:wedge}, within each municipality ($j$), the number of households ($\mathit{volume}_{i,j}$) in each income bracket is multiplied by the corresponding median income and integrated to calculate the grand total. Subsequently, this grand total is divided by the total number of households in the respective municipality to obtain the average annual income in JPY ($W_{j,\text{JPY}}$).

\begin{equation}
\label{eq:wedge}
W_{j,\text{JPY}} = \frac{\sum_{i=1}^{9} \mathit{income}_{i,j} \times \mathit{volume}_{i,j}}{\sum_{i=1}^{9} \mathit{volume}_{i,j}} \times 10,000
\end{equation}

To facilitate international comparisons, these values were converted into US Dollars (USD) and Euros (EUR) using the Bank of Japan exchange rates as of August 3, 2026 \cite{jpbank}. The resulting values are defined as the average annual income ($W_{j,\text{USD}}$ and $W_{j,\text{EUR}}$).

The correlation between nighttime lights ($L_j$) and the average income is presented in \figref{fig:income_nl_graph}, and the respective regression equations are shown in \eqref{eq:wedge_corr_usd} and \eqref{eq:wedge_corr_eur}.

\begin{equation}
\label{eq:wedge_corr_usd}
W_{j,\text{USD}} = 555.578 \ln(L_j) + 11684.3 \quad (R^2 = 0.143, p < 0.0001)
\end{equation}

\begin{equation}
\label{eq:wedge_corr_eur}
W_{j,\text{EUR}} = 482.717 \ln(L_j) + 10152.0 \quad (R^2 = 0.143, p < 0.0001)
\end{equation}

The rationale for employing a logarithmic function to model average annual income is based on the assumption that income changes with elasticity relative to nighttime light intensity. From \eqref{eq:wedge_corr_usd}, \eqref{eq:wedge_corr_eur}, and \figref{fig:income_nl_graph}, it can be concluded that while average annual income in each municipality is not entirely independent of brightness, the correlation is extremely weak ($R^2 \approx 0.143, p < 0.0001$).

The vertical intercept ($\beta_0$) in Equations~\eqref{eq:wedge_corr_usd} and \eqref{eq:wedge_corr_eur} represents the baseline average income when nighttime light intensity is at its minimum ($\ln(L_j) = 0$). The estimated baseline values of approximately 11,684~USD (10,152~EUR)\footnote{All monetary values originally recorded in JPY were converted into USD and EUR based on the Bank of Japan exchange rates as of August 3, 2026 ($1\text{ USD} = 156.50\text{ JPY}$ and $1\text{ EUR} = 180.69\text{ JPY}$) \cite{jpbank}.} correspond to the regional income floor in Japan. This reflects the structural reality that even in the least illuminated municipalities, average household income remains around this baseline.

\section{Conclusion}

This study investigated the general versatility of nighttime light (NTL) data by examining municipal-level NTL intensity alongside diverse public socio-economic indicators in Japan. The empirical findings confirm a strong positive association between NTL radiance and commercial activity, particularly in retail and commercial store density. Conversely, public infrastructure metrics—such as healthcare facilities and sewerage systems, which are policy-driven to maintain baseline regional equity—exhibit a more uniform distribution relative to NTL emissions. These contrasting patterns validate the capacity of NTL intensity to serve as a high-resolution proxy for both market-driven and publicly balanced human activities.

As highlighted in synthesis literature (e.g., Kurata \cite{kurata}; Henderson et al. \cite{henderson}), a fundamental contribution of NTL analytics is its capacity to bypass cross-border measurement inconsistencies and quantify localized economic output within subnational boundaries. Beyond this spatial continuity, this paper emphasizes an equally critical temporal advantage offered by Earth observation platforms.

Modern satellite portals, such as NASA's Worldview and JAXA's G-Portal \cite{gportal}, now distribute satellite observations within short latency periods (ranging from one day to a few weeks post-observation). By applying standardized cloud-masking and radiometrical calibration routines, near-real-time NTL imagery can capture sudden socio-economic shocks far faster than traditional administrative census reporting.

While this study relied on pre-calibrated annual composites provided by NOAA, our findings lay the groundwork for high-frequency economic impact tracking. Future research will focus on developing real-time monitoring methodologies using active sensors—such as "SHIZUKU" (GCOM-W) \cite{gcom_w} and "SHIKISAI" (GCOM-C) \cite{gcom_c} available through G-Portal—to evaluate rapid economic downturns, mobility contractions, and recovery phases during major societal crises, including the COVID-19 pandemic.

\paragraph{\texorpdfstring{Declaration of Generative AI in Scientific Writing}{Declaration of Generative AI in Scientific Writing}}
During the preparation of this work, the authors used ChatGPT-4 (OpenAI, accessed July 8, 2026) and Google Gemini (Google, accessed July 8, 2026) to enhance language clarity and readability. Specifically, these tools were used for \LaTeX{} formatting, including the embedding and placement of figures, table creation, and Bib\TeX{} reference adjustments.

\clearpage
\bibliographystyle{plainnat}
\bibliography{jssij2024_fixed_en}

\end{document}